\documentclass[10pt,preprint]{aastex}
\shorttitle{Remote M31 Major Axis Field}
\shortauthors{Rich et~al.}

\begin{document}

\title{Deep Photometry in a Remote M31 Major Axis Field Near
G1\altaffilmark{1}}

\altaffiltext{1}{Based on observations made with the NASA/ESA Hubble Space
Telescope obtained at the Space Telescope Science Institute, which is
operated by the Association of Universities for Research in Astronomy, Inc.,
under NASA Contract NAS~5-2655.  These observations were made in connection
with proposal GO-9099.}

\author{R.\ Michael Rich\email{rmr@astro.ucla.edu}
David B.\ Reitzel\email{reitzel@astro.ucla.edu}}
\affil{Department of Physics and Astronomy, University of California at Los
Angeles, Math-Sciences 8979, Los Angeles, CA 90095-1562, USA}

\author{Puragra Guhathakurta}
\affil{UCO/Lick Observatory, University of California at Santa Cruz, 1156
High Street, Santa Cruz, CA 95064, USA}
\email{raja@ucolick.org}

\author{Karl Gebhardt}
\affil{Department of Astronomy, University of Texas at Austin, C1400, Austin,
TX 78712, USA}
\email{gebhardt@astro.utexas.edu}

\and

\author{Luis C.\ Ho}
\affil{The Observatories of the Carnegie Institution of Washington, 813 Santa
Barbara Street, Pasadena, CA 91109}
\email{lho@ociw.edu}

\begin{abstract}
We present photometry from {\it Hubble Space Telescope\/} ({\it HST\/})/Wide
Field Planetary Camera~2 parallel imagery of a remote M31 field at a
projected distance of $\sim 34$ kpc from the nucleus near the SW major axis.
This field is near the globular cluster G1, and near one of the candidate
tidal plumes identified by Ferguson et~al.\ (2002).  The F606W ($V$) and
F814W ($I$) images were obtained in parallel with Space Telescope Imaging
Spectrograph spectroscopy of G1 (GO-9099) and total 7.11~hours of integration
time --- the deepest {\it HST\/} field in the outer disk of M31 to date,
reaching to $V\sim28$.  The color-magnitude diagram of the field shows a
clearly-defined red clump at $V=25.25$ and a red giant branch consistent with
$\rm[Fe/H]\approx-0.7$.  The lack of a blue horizontal branch contrasts with
other M31 halo fields, the Andromeda dwarf spheroidals, and with the nearby
globular cluster G1.  Comparing the observed luminosity function to the
Padova models, we find that at least some of the stellar population must be
younger than 6--8~Gyr.  The outermost detected neutral hydrogen disk of M31
lies only 2~kpc in projection from our field.  The finding that some giants
in the field have radial velocities close to that of the neutral hydrogen gas
(Reitzel, Guhathakurta, \& Rich 2003) leads us to conclude that our field
samples the old, low-surface-brightness disk rather than the true
Population~II spheroid or the remnants of a disrupted M31 satellite.

\end{abstract}


\section{Introduction}

In the course of our {\it Hubble Space Telescope\/} ({\it HST\/})
spectroscopic study of the M31 globular cluster G1 (GO-9099, Cycle~10), we
obtained a deep parallel Wide Field Planetary Camera~2 (WFPC2) image of a
nearby field in the halo of M31.  The globular cluster G1 is one of the most
luminous in the Local Group and has an substantial abundance spread (Meylan
et~al.\ 2001).  There is dynamical evidence for a central black hole
(Gebhardt, Rich, \& Ho 2002) based on the analysis of the Space Telescope
Imaging Spectrograph (STIS) spectra of M31 obtained in the {\it HST\/}
campaign.  The object G1 has been proposed as the nucleus of a dwarf
spheroidal galaxy perhaps in the process of being tidally disrupted in the
M31 halo.  This argument has been strengthened by the survey of Ferguson
et~al.\ (2002) which shows what appears to be a tidal plume of stars located
very near the position of G1 (at least in projection).  If the tidal plume
did in fact originate from a dwarf spheroidal galaxy similar to the
present-day Andromeda dwarfs, it might be expected to have a blue horizontal
branch (HB) in its color-magnitude diagram (CMD), a feature that is seen in
every M31 dwarf spheroidal companion (Da~Costa et~al.\ 1996, 2000; Da~Costa,
Armandroff, \& Caldwell 2002).  Alternatively, the tidal plumes might be
related to a companion similar to the compact elliptical satellite M32, which
is undergoing tidal stripping (Choi, Guhathakurta, \& Johnston 2002) and for
which existing deep CMDs show only a red HB (Grillmair et~al.\ 1996).  Deep
images in the vicinity of G1 might detect the stellar population from the
hypothetical dwarf spheroidal galaxy of which G1 is proposed to be the
nucleus.

The STIS spectroscopic exposure was long enough to permit a total of 7.1~hr
of WFPC2 integration on a parallel imaging field $5.3'$ N of G1.  The field
lies $\sim6$ (extrapolated) radial scale lengths out in the outer ``disk'' of
M31 (Walterbos \& Kennicutt 1988).  This parallel image is much longer than
typical {\it HST\/}/WFPC2 imaging of the M31 halo (typically a total of
$\sim4$~orbits in two colors), and is in fact the deepest image of a remote
M31 major axis field obtained to date.

The first deep stellar photometry of the M31 halo was undertaken by Mould \&
Kristian (1986), revealing an old red giant branch (RGB) with a wide range of
metal abundance.  A series of ground-based and {\it HST\/} studies
(e.g.,~Durrell, Harris, \& Pritchet 1994, 2001; Rich, Mighell, \& Neill
1996b; Holland, Fahlman, \& Richer 1996; Reitzel, Guhathakurta \& Gould 1998;
Sarajedini \& van Duyne 2001) find $\rm[Fe/H]\sim-0.7$ in a number of M31
halo fields.  These {\it HST\/} studies of M31 halo fields show that the HB
is dominated by the red clump with roughly 15\% blue stars (see Bellazzini
et~al.\ 2003) and that the RGB is metal rich, as mentioned earlier.

Two prior deep imaging studies sampling M31's disk are worth noting here.
Rich et~al.\ (1996a) image the globular cluster G1 using the PC1 CCD on WFPC2
and use the WF CCD frames to examine the field population very near the
location of the field presented in this paper.  They find that the field
population is metal rich ($-0.7$~dex).  This early study notes that the field
luminosity function (LF) rises anomalously steeply below the HB (relative to
an old population and relative to G1), but the photometry is too shallow to
draw a firm conclusion.  Although we do not reach the main-sequence turnoff
point, our new photometry suggests that the excess of faint stars noted by
Rich et~al.\ (1996a) is real.

Ferguson \& Johnson (2001) analyze archival WFPC2 imagery of an outer disk
field on the opposite side of the major axis, some 30~kpc NE of the nucleus.
They find a red clump and a metal-rich RGB, similar to the findings of the
Rich et~al.\ (1996a) and present studies.  Ferguson \& Johnson reach $V=27$
and do not report a LF, but argue from the luminosity and characteristics of
the red clump that the bulk of the stars must be $>8$~Gyr old.  If these
remote M31 fields are dominated by stellar debris from prior interactions,
one might expect to find differences among the CMDs of the various fields.

Recent ultra-deep imaging of a SE minor-axis field (11~kpc from M31's nucleus
in projection) using {\it HST\/}'s Advanced Camera for Surveys (Brown et~al.\
2003) reaches well below the main-sequence turnoff for an old stellar
population.  This study indicates that the most metal-rich stars, about a
third of all the stars along this line of sight, are only $\sim 6$~Gyr old.
A possible hypothesis is that these stars have been stirred up from M31's
disk.  The spectroscopic study by Reitzel \& Guhathakurta (2002) in a more
remote minor-axis field also finds that the most metal-rich giants may be
related to M31's disk, and preliminary spectroscopic results suggest that the
same may be true in the Brown et~al.\ field (Guhathakurta \& Reitzel 2002).
The outer disk of M31 and the disk-halo transition region may provide
critical clues for understanding the formation of large spiral galaxies like
our own.

\section{Observations}

The WFPC2 imaging presented in this paper was obtained in October~2001, in
parallel with STIS spectroscopy of the core of G1.  Because small moves were
executed only along the STIS slit, we do not have a true dither pattern for
our field.  Constrained by the parallel observations, our field lies at
$\rm\alpha(J2000)=00^h 32^m 46.3^s$, $\delta(J2000)=+39^\circ 40' 00.1''$,
offset from G1 only in declination by $+5.34'$.  In projection, the field
lies $\approx1.2$~kpc from G1, and $\approx34$~kpc from the nucleus of M31.
This places our field well outside the tidal radius measured by Meylan
et~al.\ (2001), yet near enough to G1 so that it should sample the population
of any putative dwarf spheroidal galaxy that might once have been associated
with it.  Our field also falls within a possible tidal plume near the M31
outer disk/halo (Ferguson et~al.\ 2002).

Our total integration time is 3.11~hr in F606W and 4.0~hr in F814W (roughly
equivalent to a broad $V$ and $I$ band, respectively).  This is a factor
of~2.7 longer than typical M31 globular cluster images, which usually total
only 4~orbits of integration time.

\section{Data Analysis}

Bias and flat-field corrections are applied to the data as part of the
standard {\it HST\/} data processing pipeline.  Masking of bad columns and
pixels, cosmic-ray removal, and masking of hot pixels are accomplished using
the utilities provided in the HSTphot package (Dolphin 2000).  The cleaned
images are aligned and median combined to produce a final F606W and F814W
image.

Stars are identified and fitted on the combined F606W+F814W image using
HSTphot.  HSTphot is a point-spread-function (PSF)-fitting stellar photometry
package similar to DAOPHOT (Stetson 1987), and is designed to handle
undersampled PSFs such as those found in WFPC2 images.  HSTphot uses a
library of TinyTim PSFs for different locations across the CCD and for
different locations of the star's center within the pixel.  PSF fits are
also carried out independently in the coadded F606W and F814W images in order
to derive stellar photometry in the two bands.  The instrumental {\it HST\/}
magnitudes are converted to Johnson/Cousins $V$ and $I$ using the color
transformation relations of Holtzman et~al.\ (1995).  We also performed
experiments using DAOPHOT and simple aperture photometry, but found no
improvement over our approach.

Objects are required to have an {\it overall\/} $\rm S/N>5$ in the combined
F606W+F814W image and a $\rm S/N>5$ in the F606W band ($V\lesssim28.1$).  The
use of such conservative thresholds ensures that the sample is practically
free of noise detections as confirmed by subsequent visual inspection.  A
further cut of $\rm S/N>3$ in the F814W band ($I\lesssim27.3$) is applied for
the purpose of constructing the CMD whereas the F814W cut is relaxed to
$1\sigma$ ($I\lesssim28.5$) for the construction of the LF.  The exact values
of these 5-, 3-, and 1-$\sigma$ limiting magnitudes depend on the ``local
sky'' background level which varies across the image because of stellar
surface density fluctuations and proximity to bright stars; the mean limiting
magnitudes for the sample are listed above.  In addition to the S/N cuts,
objects are required to pass a $\chi<1.5$ cut (i.e.,~good fit to the {\it
HST\/} PSF) and $-0.4\leq{\tt sharpness}\leq0.2$ cut (to reject noise spikes,
galaxies, and blends).  Finally, each surviving candidate star is inspected
on the coadded WFPC2 F606W and F814W images.  About 6\% of these objects are
found to be spurious (mostly diffraction spikes and saturation bleed
artifacts near bright stars) and are eliminated from the sample.

Using HSTphot, artificial stars with $20<V<30$ and $-1<V-I<4$ are inserted
into the images.  They are added in groups of~1750 over the field of each WF
CCD, and in groups of 50 for the smaller PC1 field, in ten separate runs.
For each run, the resulting image is analyzed with HSTphot and the recovered
objects are required to pass the same selection criteria as the program
stars.  To quantify the degree of completeness and error/systematic bias in
the photometry of faint stars, we construct a recovery matrix.  Its rows and
columns correspond to the input (true) and recovered magnitudes of the
artificial star, respectively, while the matrix elements give the fraction of
stars recovered at all input/output apparent magnitudes.  We multiply the
theoretical LFs by the recovery matrix, and compare the result to our
observed LF.

\section{Discussion}

Figure~1 shows the CMD for our remote M31 major-axis field near G1.  The
appearance of the CMD of the field population is consistent with that
reported by Rich et~al.\ (1996a): there is a clearly-defined RGB, while the
HB is in the form of a red clump and lacks blue stars.  Most significantly,
there appears to be a sharp rise in the number of stars fainter than the red
clump.  This feature is confirmed after careful exclusion of spurious
objects, after accounting for background galaxies, and via artificial star
tests.  The observed color spread at faint end of the CMD is mostly driven by
F814W photometric errors.  By contrast, the LF analysis described below does
not rely on F814W photometry; it takes advantage of the greater depth of the
F606W data by relaxing the F814W S/N threshold used for sample selection
(\S\,3).

Isochrones from the Padova models (Girardi et~al.\ 2000) are overlaid on the
CMD.  The red giants straddle the $\rm[Fe/H]=-0.7$ isochrone for an old
(13~Gyr) population.  The field giant metallicities inferred from a
comparison with Galactic globular cluster RGB fiducials are identical to
those derived using model isochrones.  The photometric metallicity estimates
would be $\sim0.2$~dex higher if the field population were of order 8~Gyr old
instead of 13~Gyr.  Not surprisingly, these conclusions agree with those
drawn by Rich et~al.\ (1996a) based on field stars directly adjacent to G1.
The relatively high metallicity in these outer disk regions of M31 contrasts
with the lower values found in the outer reaches of the Local Group Sc
galaxies M33 and NGC~2403 (Davidge 2003).  One may attribute this difference
to the considerably lower luminosities and later Hubble types of these
galaxies.  Our M31 field CMD is similar to that seen in the outer disk field
of Ferguson \& Johnson (2001); however, their field appears to have a more
prominent blue extension (presumably a small number of young main sequence
stars).

There is a handful of blue objects with $0\lesssim(V-I)\lesssim0.5$ and
$25\lesssim{V}\sim26$ in our G1 field CMD.  These may be compact background
field galaxies, even though they look unresolved upon inspection of the
coadded WFPC2 F606W and F814W images, and pass the $\chi$ and {\tt sharpness}
criteria for stars.  There is also the possibility that these blue objects
are young main-sequence stars in M31's disk, given the association we
are finding with neutral hydrogen in the vicinity (see \S\,4.1).

Figure~2 shows the central result of our paper: the stellar LF in the $V$
band in our 34~kpc major-axis field near G1 compared to models.  Two points
are immediately evident from this plot and the CMD in Figure~1.  First, the
HB is in the form of a red clump, indicating a large population of stars
older than a few~Gyr (assuming age is the ``second parameter'' that
determines HB morphology).  Second, a model in which the entire population is
old is not a good fit to the data.  The $V$-band LF is formed by applying a
less stringent selection criterion than was used for the CMD: $>5\sigma$ in
F606W, but only $>1\sigma$ in F814W, combined with our shape criteria and
visual inspection.  Even though this allows marginal F814W detections to be
accepted, we feel confident with this choice because: (a)~the F606W and
overall S/N criteria are conservative, (b)~the predicted galaxy counts are
relatively low (circles in Figure~2) so contamination from misclassified
galaxies should be negligible, and (c)~the nature of the CMD suggests that we
are measuring bona fide stars in M31.  While it appears that some
intermediate-age stars must be present, Figure~2 also shows that much of the
field is likely old and is not dominated by stars significantly younger than
5~Gyr.

Padova LFs (Girardi et~al.\ 2000), most with $\rm[Fe/H]=-0.7$ and one with
$\rm[Fe/H]=-1.3$, are normalized to our observed M31 field LF at $V=26$, just
below the HB.  (It is impractical to normalize at the HB in Figure~2 because
the $V$~magnitude of the HB peak varies with age/metallicity and the models
do not all match the HB peak position seen in the data.)  The theoretical LFs
have been multiplied by the artificial star results matrix.  While the
matrix-convolved theoretical LFs match the overall shape of the observed LF,
they fail to reproduce the faint end in detail: incompleteness is observed to
set in about 0.5~mag brighter than the predictions of the artificial star
tests.

The mismatch in the faint end rolloff has no effect on our finding that the
observed counts exceed, by more than a factor of three, the prediction for a
purely old stellar population.  A lower metallicity does produce a slightly
brighter turnoff at a given age (see model LF with $\rm[Fe/H]=-1.3$ shown by
the short dashed curve in Figure~2), but this does not match the color/shape
of the RGB (Figure~1).  There is no plausible alternative to normalizing the
theoretical LFs to the data that changes this result.  The observed galaxy
counts in the Hubble Deep Field South are plotted as open circles in Figure~2
and show no rise which can account for the observed LF (Volonteri et~al.\
2001).  More precise constraints on the age will require deeper data, but for
the moment, we conclude that the stellar population near G1 contains a large
number of intermediate-age stars.

The bright portions of the model LFs (RGB, HB) in Figure~2 do not agree
perfectly with the data.  The field giants in the observed LF are deficient
in number by about a factor of two at $V\sim 24$, and the HB bump at
$V\sim 25.25$ (sensitive to the He burning lifetime) is slightly underabundant
relative to the models.  The models may not be particularly well suited to
predicting these late stages of stellar evolution.  Moreover, the actual
population is almost certain to be a composite spanning a range in age and
metallicity.

The skeptical reader might choose to ignore any data fainter than $V=27$,
even though we have taken great care to ensure the validity of faint star
measurements, and despite the very clear stellar detections at yet fainter
magnitudes.  Even at $V=27$, the old population falls short by a factor
of~2.5 in explaining the observed counts.  We conclude that there is
compelling evidence for intermediate-age stars in this field, but that deeper
imaging is required to establish definitive proof of their existence and to
determine their precise age distribution.

We are convinced that the observed rise in our field LF starting at
$V\sim26.5$ is {\it not\/} due to noise spikes mismeasured as stars near the
limiting magnitude of our data.  As a check, we have reduced archival WFPC2
images of the cluster G1 (R.M.R.'s GO-5464 program) using HSTphot in a manner
identical to that described in \S\,3.  Figure~3 compares the resulting G1
cluster LF to our field LF, with the curves normalized at the HB (the LFs in
Figure~2 were normalized at $V=26$ but that is uncomfortably close to the
incompleteness limit of the G1 LF).  While the cluster dataset is 1.5~mag
shallower because of its shorter exposure time, it does not show any rise due
to spurious detections near its limiting magnitude.

\subsection{A Connection to the Outermost Disk Population?}

The presence of an enormous warped disk of neutral hydrogen (Newton \&
Emerson 1977) and an exponential distribution of carbon stars extending out
to a radius of 35~kpc (Battinelli, Demers, \& Letarte 2003) both favor the
existence of an extended disk in M31.  We argue that the underlying stellar
population in our 34~kpc major-axis field is consistent with a disk
population that is many Gyr old.

The accompanying paper by Reitzel, Guhathakurta, \& Rich (2003) reports on
spectroscopy of red giants in the G1 field.  They find that the mean radial
velocity of the main concentration of red giants, $\rm-450~km~s^{-1}$,
disagrees with both the radial velocity of G1 and the systemic velocity of
M31, but is in excellent agreement with the radial velocity of neutral
hydrogen in the disk of M31 located 32~kpc from the nucleus (Cram, Roberts,
\& Whitehurst 1980).  These observations point to a kinematic link between
the stellar population near G1 and the neutral hydrogen in the outer disk of
M31.  We are led to conclude that the faint stellar population near G1
relates to the extended formation history of the outer disk and does not
belong to a dwarf spheroidal galaxy associated with G1, as some have
suggested.

Is this population actually the thick disk of M31 proposed by Sarajedini \&
van Duyne (2001)?  The CMD in Figure~1 and especially the LF in Figure~2 are
consistent with an intermediate-age population, while the Milky Way thick
disk is considered to be old.  We would further argue that the kinematic
connection to the neutral hydrogen lying only some 2~kpc away (in projection)
would not be expected for a thick disk population resembling that of the
Milky Way (old, metal poor, and with kinematics intermediate between those of
the thin disk and halo).  Moreover, the velocity dispersion of the putative
disk stars in the vicinity of G1 is too low to be consistent with a thick
disk origin (Reitzel et~al.\ 2003).  Deeper photometry and a larger sample of
radial velocities will be required to settle the reality of this connection.

The CMD in our major-axis field bears little resemblance to those of the
Andromeda dwarf spheroidal companions, which always have some blue HB stars
and are usually more metal poor (Da~Costa et~al.\ 1996, 2000, 2002).  There
is also no similarity to the outer fields of M32 (Grillmair et~al.\ 1996)
where the giants are much more metal rich ($\rm[Fe/H]=-0.4$).  Unlike our
field, G1 has a well-detected blue HB, but the RGB of G1 is similar to that
of our field.  As emphasized in the earlier study of Rich et~al.\ (1996a),
the field population near G1 most closely resembles other M31 halo fields.
An {\it HST\/} field on the opposite side of the major axis of M31 has been
analyzed (Ferguson \& Johnson 2001; Bellazzini et~al.\ 2003) and
qualitatively resembles our field, though there is a more significant hint of
a blue HB in that field.

Our CMD more closely resembles the bright portion of the ultra-deep
minor-axis halo field studied by Brown et~al.\ (2003).  A metal-rich RGB
dominates the population, and some fraction of the stars in our field appear
to be youthful.  Were it not for the kinematic connection to the outer
neutral hydrogen disk of M31, it would be difficult to argue that the stellar
population is not somehow connected to the halo field of Brown et~al.\
(2003).  We conclude that the spatial location, kinematics of the bright
giants, and the extreme sparseness of the blue HB all favor the connection to
the outer disk rather than to the overall halo, but only deeper photometry in
this field can settle the issue.

\section{Conclusions}

We present photometry for a remote major-axis field of M31 located $5.34'$ N
of the luminous globular cluster G1.  The total integration time is 7.11~hr
and the photometry reaches $V\sim28$, making this one of the deepest fields
imaged in the M31 outskirts, aside from the recent Brown et~al.\ (2003) study
of a halo field.  There is a clearly-defined RGB with $\rm[Fe/H]\approx-0.7$;
the HB is represented by the red clump only.  We conclude that the steep rise
seen near the faint end of the LF must be due to the onset of the subgiant
branch in an intermediate-age stellar population.  Noting the agreement
between the stellar kinematics in this field and that of the neutral hydrogen
in the outer disk of M31 (Reitzel et~al.\ 2003), we are led to suspect that,
in addition to what appears to have been an old star-formation episode, the
outer disk gas has continued to fuel more recent star formation.

We conclude that, in general, the stellar population in the field near G1 is
different from that of the cluster itself.  While their RGBs and red clumps
are comparable, the field lacks a blue HB, even with our deeper photometry.
This lack of a blue HB also distinguishes the field from the Andromeda dwarf
spheroidal galaxies.  The qualitative properties of the CMD in our major-axis
field resemble those seen in other M31 halo fields.  However, the field star
LF is different from those of old metal-rich globular clusters like 47~Tuc in
the Milky Way or G1 in M31.

The LF shows evidence of an extended and complex star formation history and a
connection to the outer neutral hydrogen disk of M31, as supported by the
dynamics reported in Reitzel et~al.\ (2003).  The dominant population in this
field is $>5$~Gyr old, leading us to conclude that the outer disk of M31 had
a significantly greater rate of star formation in the past than it does
today.  A similar conclusion is reached by Ferguson \& Johnson (2001).
Despite our field lying in a relatively disturbed region of the M31 halo, the
CMD is remarkably similar to that found by Ferguson \& Johnson on the
opposite side of the M31 disk.  The high metallicity seen in both of these
outer disk fields is interesting, as the stars have formed in a very
low density environment.

The inferred higher rate of star formation $>5$~Gyr ago would imply a
brighter M31 disk at a look-back time of a third of the Hubble time, broadly
consistent with what is seen in disk galaxies at moderate redshift (Lilly
et~al.\ 1998).  Rotation curve measurements of field and cluster spirals
indicate that the Tully-Fisher (TF) relation for large galaxies like the
Milky Way and M31 was more or less in place by $z\sim1$, albeit shifted with
respect to the local relation suggesting that galaxies were brighter by
about 1~mag in the rest-frame optical 7~Gyr ago (e.g.,~DEEP1 survey---Vogt
et~al.\ 1996).  This is not necessarily in conflict with our finding of a
$\sim4$~Gyr population in M31's outer disk.  Firstly, not all the stars in
the G1 field are of intermediate age; there probably is an underlying
population of stars older than 7~Gyr and it is presumably their counterparts
that are being sampled by the $z\sim1$ TF surveys.  Secondly, our G1 field is
located in M31's faint outer disk whereas the distant galaxy surveys are only
sensitive to the bright star-forming inner parts of disks.

Theories of galaxy formation have met with limited success so far when it
comes to understanding the disk properties of the Milky Way and M31 or the TF
relation in general---the so-called angular momentum crisis in models
(cf.~Klypin, Zhao, \& Somerville 2002).  One hopes that the combination of
``direct look back'' studies of distant spiral galaxies (such as the ongoing
DEEP2 survey) and the application of ``fossil-record'' techniques to nearby
spirals (as in our study) will ultimately produce a comprehensive picture of
the formation and evolution of disk galaxies.

\acknowledgments

Support for proposal GO-9099 was provided by NASA through a grant from the
Space Telescope Science Institute, which is operated by the Association of
Universities for Research in Astronomy under NASA contract NAS~5-26555.
These observations are associated with proposal GO-9099.  We thank Leo
Girardi for providing the model luminosity functions.  We thank Marla Geha
and Annette Ferguson for a careful reading of and helpful comments on the
manuscript.  P.G.\ acknowledges support from the National Research Council of
Canada in the form of a 2002-2003 Herzberg fellowship, and is grateful to the
Herzberg Institute of Astrophysics for hosting him during that time.  He
would also like to thank Joe Miller for his support through a UCO/Lick
Observatory Director's grant.

\clearpage

\begin{figure}
\plotone{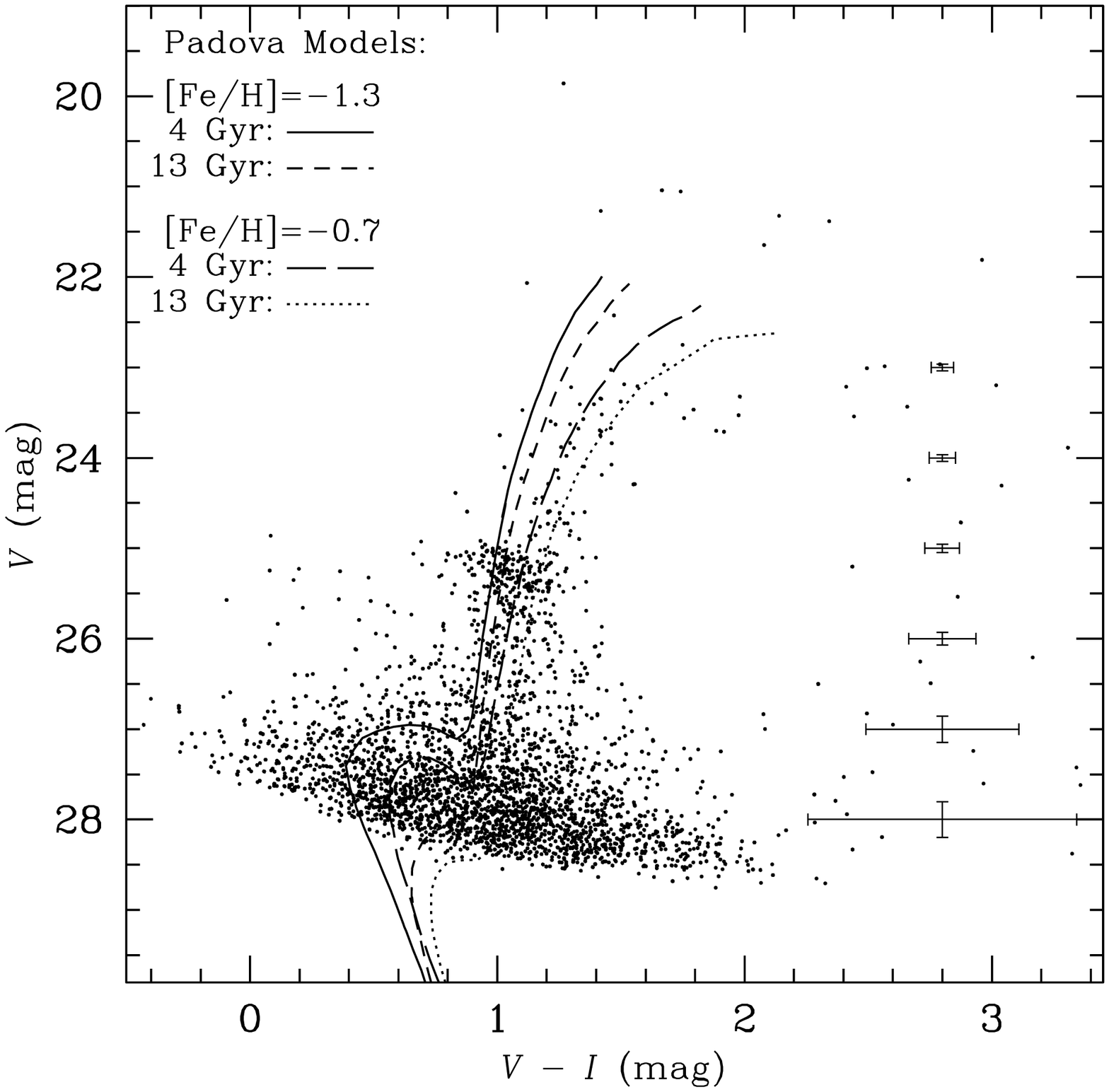}
\figcaption[Rich.fig01.eps]{Color-magnitude diagram of the field population
near G1, based on  detection cuts of $5\sigma$ in F606W and $3\sigma$ in
F814W (along with the other selection criteria listed in \S\,3).  The
$\pm1\sigma$ photometric errors in magnitude and color are indicated as a
function of $V$ magnitude.  Overplotted are 4 and 13~Gyr isochrones for each
of $\rm[Fe/H=-1.3$ and $-0.7$ from the Padova models (Girardi et~al.\ 2000);
the G1 field population is consistent with $\rm[Fe/H]\approx-0.7$.  The red
clump and giant branch are well detected.  Notice the large number of stars
at the faint end ($V\sim27$).  Visual inspection of the coadded WFPC2 images
indicates that these are real star-like objects.  The same is true for the
blue objects at $V-I\sim0.3$ and $V\sim25.5$; these may be unresolved compact
background field galaxies or possibly young main-sequence stars in M31's
disk.}
\end{figure}

\clearpage

\begin{figure}
\plotone{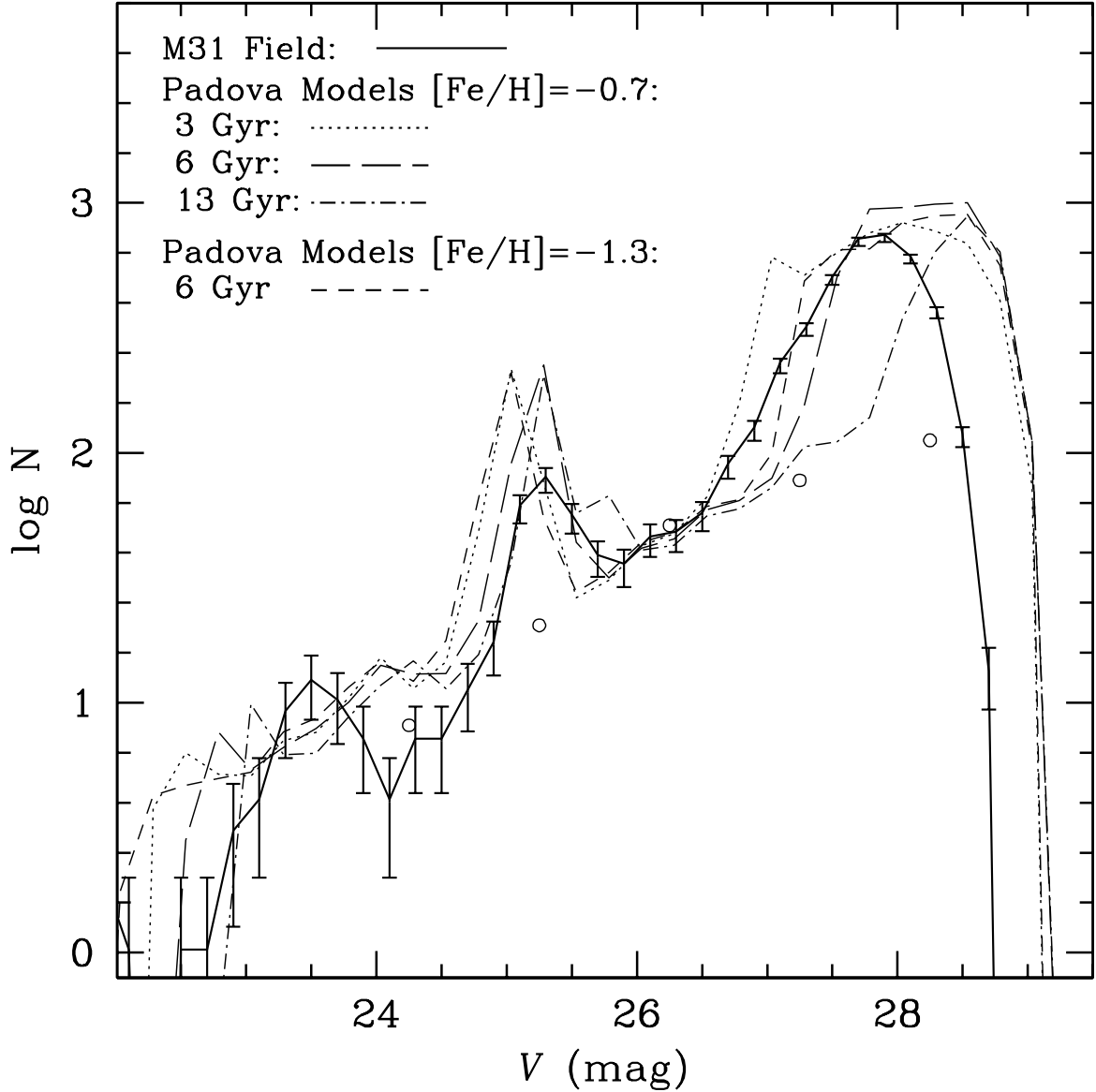}
\figcaption[Rich.fig02.eps]{Luminosity function of stars in our G1 field
(solid line), compared with model LFs from the Padova isochrones (Girardi
et~al.\ 2000) with age $t=3$, 6, and 13~Gyr and $\rm[Fe/H=-0.7$, and
$t=6$~Gyr and $\rm[Fe/H=-1.3$.  All models have been convolved with our
artificial star recovery matrix (see \S\,3) and normalized to match the data
at $V=26$.  For this plot, we accept stars detected above $5\sigma$ in F606W
and above $1\sigma$ in F814W; the rationale for relaxing the $I$-band
detection requirement is discussed in \S\,4.  The open circles show faint
galaxy counts from the Hubble Deep Field South (Volonteri et~al.\ 2001).
Notice the significant difference between the dot-dashed line for the 13~Gyr
old population and our counts.  We conclude that the LF is not consistent
with a pure old stellar population.}
\end{figure}

\clearpage

\begin{figure}
\plotone{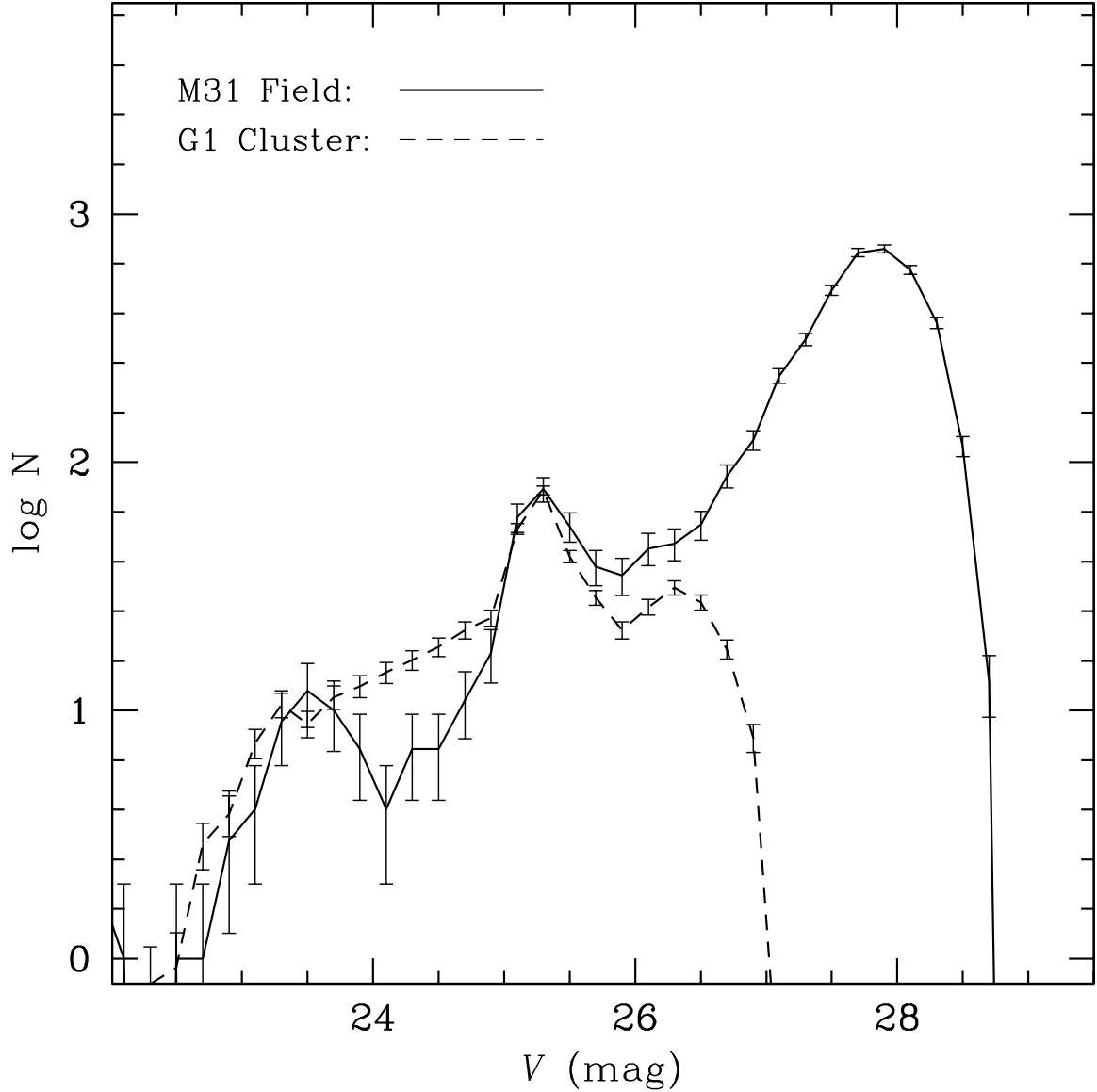}
\figcaption[Rich.fig03.eps]{Luminosity function of our deep field in the
vicinity of G1 (solid line), compared with that for the population of the G1
cluster itself (dashed line) based on data from GO-5464, reduced in the
identical way as our deep field.  The LFs have been normalized at the
horizontal branch; notice the excellent agreement in the $V$ magnitudes of
their HB peaks.  The G1 cluster data are $\sim1.5$~mag shallower than the
deep field data but nevertheless provide a sanity check: if the steep rise at
the faint end of the field LF were due to mismeasurement of noise near the
limiting magnitude of the data, a similar rise would be seen at $V\sim26$ in
the cluster LF.  There is a significant deficiency of bright red giants in
our field relative to the G1 cluster, and this may reflect shorter
evolutionary lifetimes for younger giants.}
\end{figure}

\end{document}